\documentclass{emulateapj}

\shorttitle{Galaxy Groups in the SDSS-DR3}
\shortauthors{Merch\'an and Zandivarez}

\begin{document}

\def\mpc{h^{-1} {\rm{Mpc}}}
\def\up{h^{-3} {\rm{Mpc^3}}}
\def\uk{h {\rm{Mpc^{-1}}}}
\def\lsim{\mathrel{\hbox{\rlap{\hbox{\lower4pt\hbox{$\sim$}}}\hbox{$<$}}}}
\def\gsim{\mathrel{\hbox{\rlap{\hbox{\lower4pt\hbox{$\sim$}}}\hbox{$>$}}}}

\title{Galaxy Groups in the Third Data Release of the SDSS}

\author{Manuel E. Merch\'an\altaffilmark{1}, Ariel Zandivarez\altaffilmark{1}}
\affil{Grupo de Investigaciones en Astronom\'{\i}a Te\'orica y Experimental,
IATE, Observatorio Astron\'omico, Laprida 854, C\'ordoba, Argentina.}

\altaffiltext{1}{Consejo de Investigaciones Cient\'{\i}ficas y T\'ecnicas de la Rep\'ublica Argentina.}

\begin{abstract}
We present a new sample of galaxy groups identified in the Sloan Digital 
Sky Survey Data Release 3. Following previous works we use the well tested 
friend-of-friend algorithm developed by Huchra \& Geller which take into 
account the number density variation due to the apparent magnitude limit of 
the galaxy catalog. To improve the identification we 
implement a procedure to avoid the artificial merging of small systems in high 
density regions and then apply an iterative method to
recompute the group centers position.
As a result we obtain a new catalog with 10864 galaxy groups 
with at least four members. The final group sample has a mean redshift 
of 0.1 and a median velocity dispersion of $230 \ km \ s^{-1}$.
\end{abstract}

\keywords{galaxies: clusters : general}

\section{Introduction} 
Galaxy group samples have become a very important issue in cosmology.
Since hierarchical clustering drives the structure formation in the
universe, galaxy groups can be considered a fundamental piece of the chain
joining galaxies and clusters of galaxies. 

The information obtained from these
systems allows us to understand what the internal processes
ruling the intragroup medium are and what the most suitable cosmological
model to describe their distribution is.

The first large samples of groups comprised around $\sim1000$ 
galaxy groups and were constructed from different redshift surveys
(Merch\'an, Maia \& Lambas 2000, Giuricin et al 2000, Tucker et al. 2000,
Ramella et al. 2002). 
More recently larger galaxy group catalogs were constructed from 
different releases of the 2dF Galaxy Redshift Survey (hereafter, 2dFGRS). 
The first one was constructed from the 2dFGRS 100K Data Release by 
Merch\'an \& Zandivarez (2002) with a total of $\sim2200$ groups while
using the final release of the 2dFGRS, Eke et al. (2004a)
constructed a catalog containing $\sim7000$ galaxy groups with at least
four members. 
Research involving these samples comprise studies ranging from local physical 
properties (Mart\'{\i}nez et al. 2002, Dom\'{\i}nguez et al. 2002,
D\'{\i}az et al. 2004, Eke et al. 2004b, Ragone et al. 2004) to large 
scale structure (Zandivarez, Merch\'an \& Padilla 2003, Padilla et al. 2004).

At the present, the largest galaxy redshift survey is the Third Data
Release of the Sloan Digital Sky Survey (hereafter, SDSS-DR3).
This release has approximately $\sim530000$ spectra of
which $\sim380000$ are galaxies with a redshift accuracy of $30 \ km \ 
s^{-1}$. 
The size and deepness of this sample, turns it an ideal source of information
to obtain a new galaxy group sample. 

The aim of this work is to construct a group catalog from the
galaxies in the Third Data Release of the SDSS. 
The group identification is performed using the algorithm developed by
Huchra \& Geller (1982).
We also apply a technique to improve the identification for groups
with at least ten members. 

A description of the group identification algorithm is given in Section 2 
while the identification on galaxies and the subsequent improvements are
carried out in section 3. Finally, in section 4 we summarize our results.  

\section{The group-finding algorithm}
Group identification is performed using the friend-of-friend algorithm 
developed by Huchra \& Geller (1982).
According to these authors, given a pair of galaxies with mean radial 
comoving distance $D=(D_1+D_2)/2$ and angular separation $\theta_{12}$, 
the algorithm links galaxies satisfying the following conditions:
\begin{equation}
D_{12}=2 \sin\left(\frac{\theta_{12}}{2}\right)D \leq D_L = D_0 R
\end{equation}
and 
\begin{equation}
V_{12}=|V_1-V_2| \leq V_L=V_0 R
\end{equation}
where $D_{12}$ is the projected distance and $V_{12}$ is the line-of-sight
velocity difference. Comoving distances $D_i$ are estimated using the 
relation corresponding to an Einstein-de Sitter model 
\begin{equation}
D_i(z)=\frac{c}{H_0}\int_0^z \frac{dz'}{\Omega_M (1+z')^3+\Omega_{\Lambda}}
\end{equation}
with density parameters $\Omega_M=0.3$, $\Omega_{\Lambda}=0.7$ and 
$H_0=100 h \ km \ s^{-1} \ Mpc^{-1}$.
The transverse ($D_L $) and radial ($V_L$) linking lengths
scale with $R$, to compensate for the number density variation due to 
the apparent magnitude limit of the survey.
The scaling factor $R$ is computed using the galaxy luminosity function of the
sample $\phi(M)$:
\begin{equation}
R = \left[\frac{\int_{-\infty}^{M_{12}}\phi(M)dM}{\int_{-\infty}^{M_{lim}}\phi(M)dM}\right]^{-1/3} 
\end{equation}
where $M_{lim}$ and $M_{12}$ are the absolute magnitude of the brightest 
galaxy visible at the fiducial $D_f$ and mean galaxy $D$ distances,
 respectively.
Usually $D_0$ is chosen in order to obtain the desired overdensity 
$\delta\rho/\rho$, which is given by
\begin{equation}
\frac{\delta \rho}{\rho}=\frac{3}{4\pi D_0^3}\left(\int_{-\infty}^{M_{lim}}\phi(M)dM\right)^{-1}-1
\end{equation}
and correspond to a fixed overdensity contour surrounding a group.
\begin{figure*}
\epsscale{0.85}
\plotone{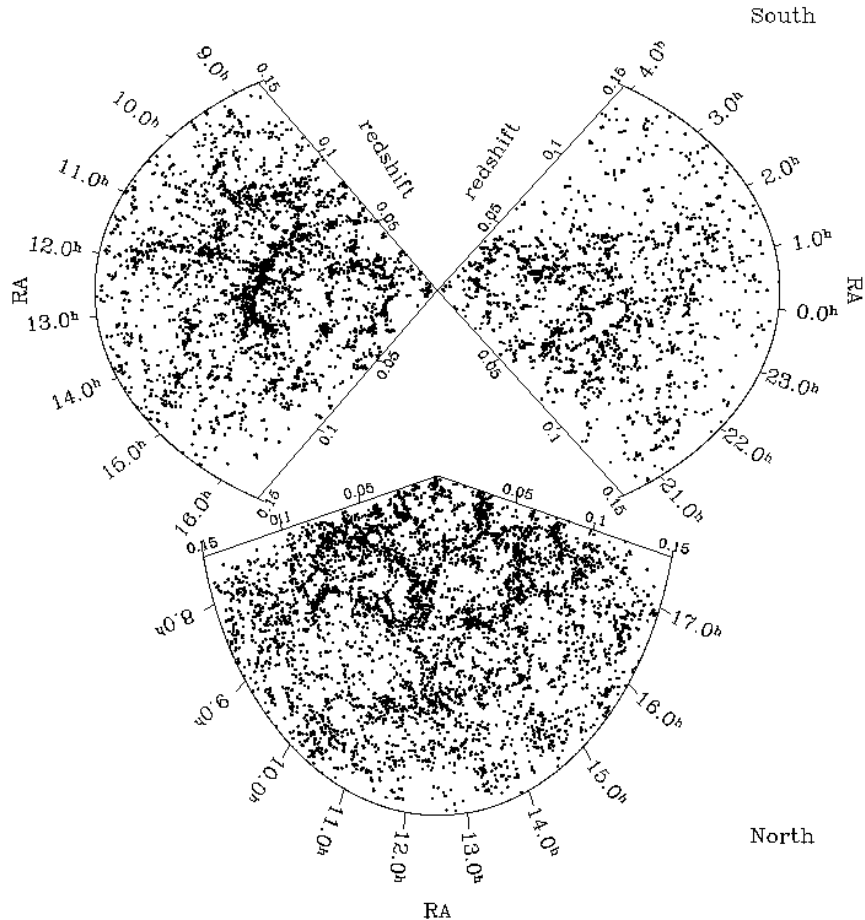}
\caption{
Pieplots showing the galaxy groups in the SDSS-DR3. The 
galaxy groups distribution is showed in three different ranges of right
ascension and for redshifts $z\leq0.15$.
\label{fig1}
} 
\end{figure*}

\section{Galaxy groups sample}
\subsection{The galaxy catalog}
The Sloan Digital Sky Survey has validated and made publicly
available the Third Data Release (Abazajian et al. 2004). 
This catalog is a photometric and spectroscopic survey covering
$4188 \ deg^2$ 
with five-band ($u \ g \ r \ i \ z$) imaging data and 528640
spectra of galaxies, quasars and stars. 
In this work we use the main spectroscopic galaxy sample which
comprises  $\sim 300000$ galaxies with redshifts
$z \leq 0.3$ and an upper apparent magnitude limit of 17.77 in the r-band.

\subsection{Group Identification}
The adopted linking lengths values are motivated by the 
group identification analysis performed by 
Merch\'an \& Zandivarez (2002) for the 2dFGRS. 
Using mock catalogs, they explore a wide range of linking lengths values
in order to maximize the group identification accuracy.
Since the SDSS-DR3 and 2dFGRS have similar redshift distributions and 
luminosity functions (Norberg et al. 2002),
we choose identical values for the group identification:
a transversal linking length corresponding to an overdensity of 
$\delta \rho/\rho=80$, a line-of-sight linking length of 
$V_0=200 \ km \ s^{-1}$ and a fiducial distance $D_f=10 \ h^{-1} Mpc $. 
The scaling factor $R$ is estimated using a galaxy luminosity function
fitted  using a Schecter function with parameters ($\alpha=-1.05\pm 0.01,
M_{\ast}-5\log{h}=-20.44\pm 0.01$) given by Blanton et al. (2003).
The main properties of the obtained groups sample
are summarized in Table \ref{tab1} (SDSS-DR3 First identification).

\subsection{Rich groups identification improvement}
\begin{figure}
\epsscale{0.85}
\plotone{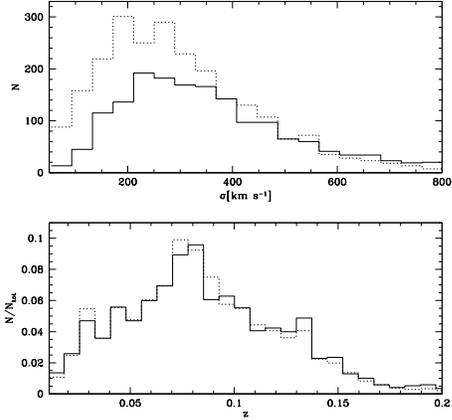}
\caption{
Velocity dispersion (upper panel) and redshift (lower panel)
distributions for rich groups. The solid line are the distributions
for groups with at least ten members after the first identification
while the dotted line shows the group distributions when the corrections 
on the identification has been applied.
\label{fig2}
} 
\end{figure}

It should be taken into account that groups obtained from galaxy redshift 
surveys have unavoidable contamination problems. 
By instance, the method described previously, can not fully avoid the 
interloper effect (i.e. spurious inclusion of non-member galaxies),
so an artificial merging of small groups with large systems is
likely to happen. 
Recently, working with groups identified in the 2dFGRS Final Release 
and the First Release of the SDSS, D\'{\i}az et al. (2004) have developed 
a method to minimize this problem on group identification. 
This method propose a second identification on galaxy groups with at least 
ten members in order to split merged systems or eliminate spurious member 
detections. 
Analyzing group identifications on mock catalogs in real and redshift 
space, they found that a higher value for $\delta \rho/\rho$ ($\sim315$)  
produce a more reliable group identification. 
Performing a redshift space identification with this density contrast
is equivalent to a group identification in real space corresponding to
$\delta \rho/\rho=80$.
Furthermore, as shown by M\'erchan \& Zandivarez (2002), adopting lower 
transversal linking length values, inproves the identification accuracy 
(see Figures 2 and 3 therein).
Therefore, following the suggestion of D\'{\i}az et al. (2004), we 
perform a second identification for groups with at least ten members.

A very accurate group center determination is required for some studies
related with the galaxy distribution inside groups.
Trying to measure density profiles of galaxy groups, D\'{\i}az et al. (2004) 
have also proposed a method for correcting group center position for those 
systems with at least ten members. 
Firstly, their method define a new group center estimator by using the  
projected local number density of each member galaxy for weighting their 
group center distances. 
The second part of the method is an iterative procedure to improve the 
group center location by removing galaxies beyond a given distance and 
recomputing the center position. 
The procedure follows until the center location remains unchanged (for more
details see section 2 of D\'{\i}az et al. 2004). 
Given that the described procedure needs to compute the projected local 
number density with five galaxies, this method only can be applied 
to groups with at least ten members.
After applying this correction, the full catalog comprise 9703 galaxy 
groups with at least 4 members.
Figure \ref{fig1} shows the spatial distribution of groups identified 
in the SDSS-DR3 with at least 4 members and $z \leq 0.15$. 

\subsection{Group physical properties}
\begin{figure}
\epsscale{0.85}
\plotone{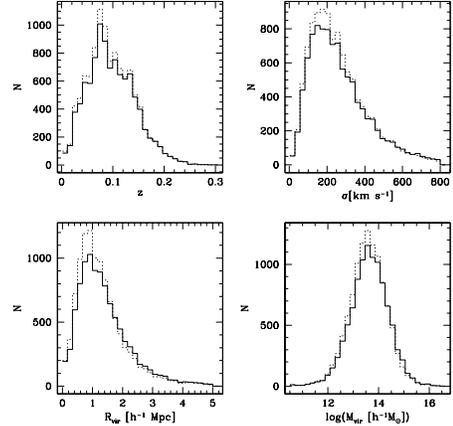}
\caption{
Histograms of the redshift (upper-left panel), velocity dispersion 
(upper-right panel), virial radius (lower-left panel) and virial mass 
(lower-right panel) distributions of groups in the SDSS-DR3. Solid lines
represent the histograms for galaxy groups after the first identification, 
while dotted lines show the histograms for the galaxy group sample after
improving the rich groups identification.
\label{fig3}
} 
\end{figure}

In addition to the group identification, we also estimate its basic physical 
properties such as velocity dispersion, virial radius and virial mass.

The line-of-sight velocity dispersion $\sigma_{v}$,
is estimated using the methods described by
Beers, Flynn \& Gebhardt (1990).
The biweight estimator is applied for groups
with richness $N_{tot}\ge 15$ whereas the gapper estimator is applied to
poor groups (Girardi et al. 1993, Girardi \& Giuricin 2000). Following 
Eke et al. (2004) we also introduce a correction due to the error in the 
redshift measurement. This is a second order correction since the 
redshift measurement error in the SDSS is $30 \ km \ s^{-1}$.\\
The virial radius is estimated using the following equation
\begin{eqnarray}
R_V & = & \frac{\pi}{2}\frac{N_g(N_g-1)}{\sum_{i>j}R_{ij}^{-1}}
\end{eqnarray}
where $N_g$ is the number of galaxy members and $R_{ij}$ the galaxy 
projected distances.\\
Finally the virial masses of galaxy groups is computed as 
\begin{equation}
M_{V}=\frac{3\sigma_{v}^2 R_V}{G}
\end{equation}
where $G$ is the gravitational constant. 

Individual group velocity dispersion estimation, allow us to observe 
rich group reideintification and recentering consequences. 
Figure \ref{fig2}
illustrate the rich group ($N_g \ge 10 $) velocity dispersion behavior 
(upper panel) for groups before and after correction (dotted and solid lines,
respectively).
As can be seen, the distribution of groups with high
velocity dispersions ($\sigma_v \gsim 300 \ km \ s^{-1}$) remains unchanged,
whereas artificially merged groups in the first identification, now appears as 
a new population with low velocity dispersion.
The redshift distribution for groups with at least ten members is shown
in the lower panel of Figure \ref{fig2}, 
It should be noted that the correction applied to
rich groups do not introduce any bias with redshift.
Histograms showing the distribution of redshift, velocity dispersion,
virial radius and virial mass of groups with at least 4 members are plotted 
in Figure \ref{fig3}.
Finally the median physical properties of the full group catalog
are quoted in Table \ref{tab1}.

\section{CONCLUSIONS}
We present the largest galaxy groups catalog at the present constructed from
galaxies in the Third Data Release of the Sloan Digital Sky Survey.
The group identification is carried out using the friend-of-friend algorithm
developed by Huchra \& Geller (1982), plus the introduction of corrections for 
rich groups which improve the group identification and the group center 
location. 
The resulting catalog comprise 10864 galaxy groups with 
at least four members and it has a mean redshift of $0.1$. 
The median basic 
physical properties of our catalog are very similar to those obtained 
in previous works (Table \ref{tab1}).

The complete galaxy group catalog with the physical properties estimated in
this work is available in the World Wide Web at 
{\it https://www.iate.oac.uncor.edu/SDSSDR3GG/} 
or upon request at any of the following e-mail addresses: {\it manuel@oac.
uncor.edu, arielz@oac.uncor.edu}.
 
\acknowledgments

This work has been partially supported by Consejo de Investigaciones 
Cient\'{\i}ficas y T\'ecnicas de la Rep\'ublica Argentina (CONICET), the
Secretar\'{\i}a de Ciencia y T\'ecnica de la Universidad Nacional de C\'ordoba
(SeCyT) and Fundaci\'on Antorchas, Argentina.

\begin{table*}
\footnotesize
\begin{center}
\caption{Median physical properties of groups}
\label{tab1}
\vskip 0.5cm
\begin {tabular}{lcccccc}
\tableline 
Sample & $N$ & $N_{gal}$ & $\bar{z}$ & $\bar{\sigma}_v(km \ s^{-1})$ & $\bar{M}(h^{-1} M_{\odot})$ & $\bar{R}_V(h^{-1}Mpc)$\\
\tableline 
  USGC (Ramella et al., 2002) & $1168$ & $6846$ & --- & $264$ & $4.7\times 10^{13}$ & $1.06$\\
  LCRS Loose Groups (Tucker et al., 2002) & $1495$ & $\sim 9250$ & --- & $164$ & $1.9\times 10^{13}$ & $\sim 1.2$ \\
 2dFGRS (Mech\'an \& Zandivarez, 2002) & $2209$ & $14634$ & $0.1$ & $232$ & $3.7\times 10^{13}$ & $1.02$\\
 2dFGRS (Eke et al., 2004a) & $7020$ & $55753$ & $0.11$ & $260$ & $3.3\times 10^{13}$ & ---\\
 SDSS-DR3 First identification & $10152$ & $78410$ & $0.1$ & $240$ & $4.2\times 10^{13}$ & $1.21$\\
 SDSS-DR3 Improved identification & $10864$ & $66517$ & $0.1$ & $230$ & $3.8\times 10^{13}$ & $1.12$\\
\tableline
\end{tabular}
\end{center}
\end{table*}

\end{document}